\documentclass[letter]{aa}
\usepackage{txfonts}
\usepackage{epsfig}
\usepackage{subfigure}
\usepackage{multirow}

\usepackage{times}
\usepackage{natbib}
\bibpunct{(}{)}{;}{a}{}{,}

\begin{document}

\title{On the Iwasawa-Taniguchi effect of radio-quiet AGN}
\author{Stefano Bianchi\inst{1,2}, Matteo Guainazzi\inst{2}, Giorgio Matt\inst{1}, Nuria Fonseca Bonilla\inst{2}}

\offprints{Stefano Bianchi\\ \email{bianchi@fis.uniroma3.it}}

\institute{Dipartimento di Fisica, Universit\`a degli Studi Roma Tre, via della Vasca Navale 84, 00146 Roma, Italy
\and XMM-Newton Science Operations Center, European Space Astronomy Center, ESA, Apartado 50727, E-28080 Madrid, Spain}

\date{Received / Accepted}

\authorrunning{S. Bianchi et al.}

\abstract
{}
{The existence of an anti-correlation between the Equivalent Width (EW) of the neutral narrow core of the iron K$\alpha$ emission line and the 2--10 keV luminosity (the so-called `X-ray Baldwin' or `Iwasawa-Taniguchi' effect) has been debated in the last years. We aim at testing this claim on the largest catalogue of radio quiet AGN high-quality X-ray spectra ever published.}
{The final sample comprises 157 objects. We search for a relation of the iron line EW not only with the X-ray luminosity, but also with the Black Hole mass, the Eddington ratio and the cosmological distance. The data presented here were analyzed homogeneously, all spectra are from the same instrument and with high Signal-to-Noise Ratio.}
{A linear censored fit on the EW versus 2--10 keV luminosity is highly significant and yields $\log(EW_{Fe}) = \left(1.73\pm0.03\right) + \left(-0.17\pm0.03\right) \log(L_{X,44})$, where $EW_{Fe}$ is the EW of the neutral iron K$\alpha$ line in eV and $L_{x,44}$ is the 2-10 keV X-ray luminosity in units of $10^{44}$ erg s$^{-1}$. The anti-correlation with the Eddington ratio is also very significant, while no dependence of the iron EW on the BH mass is apparent.}
{}


\maketitle

\section{Introduction}

\citet{baldwin77} first reported a significant anti-correlation between the EW of the [{C\,\textsc{iv}}] $\lambda 1549$ \AA line and the UV luminosity in quasars. The so-called `Baldwin-effect' was later on confirmed on more robust basis \citep[see e.g.][]{krk90} and similar anti-correlations with luminosity were found for other emission lines produced in the Broad Line Region (BLR) \citep[e.g.][]{die02}. However, the physical explanation for this phenomenon is still unclear and several possibilities are generally invoked in literature, including a change of the ionizing continuum and gas metallicity with luminosity \citep{kbf98} or a luminosity-dependent covering factor and ionization parameter of the BLR \citep{mf84}. On the other hand, some studies claim that the primary physical parameter which drives the Baldwin effect may be the accretion rate, instead of the luminosity.

In \citeyear{it93}, \citeauthor{it93} presented a similar anti-correlation between the iron K$\alpha$ emission line and the 2-10 keV luminosity, according to \textit{Ginga} observations of 37 AGN. This `Iwasawa-Taniguchi effect' (`IT effect' from now on, also known as `X-ray Baldwin effect') was then found in a larger sample of objects observed by XMM-\textit{Newton}, giving a relation between luminosity and Equivalent Width (EW) of the narrow core of the Fe K$\alpha$ as $EW\propto L^{-0.17\pm0.08}$ \citep{page04}. The significance of this effect was questioned by \citet{jb05} in their analysis of XMM-\textit{Newton} data of PG quasars, pointing out the importance of contamination from radio-loud objects. Recently, \citet{jww06} combined \textit{Chandra} and XMM-\textit{Newton} data and suggested that, after excluding radio-loud objects, the anti-correlation can be attributed to variations of the continuum, while the iron line stays constant, as expected if the two components are produced on different scales. \citet{nan97b} also reported an IT effect in their \textit{ASCA} sample of AGN, but concluded that the effect was entirely due to the relativistically broadened component of the iron line. The presence of an anti-correlation between the iron relativistic line and the luminosity was also recently found by \citet{gbd06} on a large data set of XMM-\textit{Newton} spectra of AGN.

In this paper, we search for the IT effect in the largest catalogue of radio-quiet AGN observed by XMM-\textit{Newton} published so far. Radio-loud objects are excluded on the basis of the radio-loudness parameters \textit{R} and $R_X$, calculated for each source from data in the literature. The data presented here were analyzed homogeneously, all spectra are from the same instrument and with high Signal-to-Noise Ratio (SNR).  

\section{The catalogue}

\subsection{X-ray data reduction and analysis}

Our catalogue consists of all radio-quiet Type 1 AGN observed as a main target by XMM-\textit{Newton}, whose data are public as of March 2007. Only EPIC pn \citep{struder01} data were reprocessed, with SASv7.0 \citep{sas610}. For the observations performed in Small Window mode, background spectra were generated using black-field event lists, according to the procedure presented in \citet{rp03}. In all other cases, background spectra were extracted from source-free regions close to the target in the observation event file.

Source extraction radii and screening for intervals of flaring particle background were performed via an iterative process which leads to a maximization of the SNR, similarly to what described in \citet{pico04}. Spectra were binned in order to oversample the intrinsic instrumental energy resolution by a factor $\ge3$. Moreover, spectra were further re-binned in order to have spectral bins with at least 25 background-subtracted counts, to ensure the applicability of the $\chi^2$ statistics.

We applied a number of criteria to filter the catalogue. First of all, spectra with less than 200 counts in either of the (rest-frame) bands of 0.5-2 and 2-10 keV were rejected, because they do not possess enough independent bins to be fitted with our models. Moreover, spectra affected by a pileup larger than 1$\%$ were rejected. Only two objects, namely NGC~2992 and ESO548-G081, did not have an alternative pileup-free observation and were therefore excluded from the catalogue. After the fitting procedure, all sources with a local column density (measured in the 2-10 keV band) larger than $2\times10^{22}$ cm$^{-2}$ were further excluded from the catalogue. The exclusion of radio-loud objects is crucial for the analysis of this paper. Therefore, we collected radio data at 6 and 20 cm from the literature and calculated the radio-loudness parameter \textit{R} \citep{sto92} and the X-ray radio-loudness parameter $R_X$ \citep{tw03}. According to the standard definition, all quasars with $\log(R)>1$ were excluded, while for Seyfert galaxies, known to be on average `radio-louder', $\log(R)>2.4$ and $\log(R_X)>-2.755$ were used as boundaries  \citep[see][]{panessa07}.

At the end of this selection procedure, the total catalogue comprises 157 radio quiet AGN, whose X-ray data are complemented by H$\beta$ Full Width Half Maximum (FWHM) and Black Hole (BH) mass measures, when available in literature (the coverage is 64\% and 52\%, respectively). We defer the reader to Bianchi et al., in preparation, for the full presentation of the catalogue. When more than one observation of the same object was available, we used the one with the longest exposure, independently from its flux status, in order to avoid, as much as possible, any bias for the choice.
All spectra were analyzed with \textsc{Xspec} 12.2.1 \citep{xspec}. For the aims of this paper, we restricted the analysis to the 2-10 keV band, adopting a baseline model consisting of a powerlaw plus three narrow ($\sigma\equiv0$) Gaussian lines with energies fixed at 6.4, 6.7 and 6.96 keV, corresponding to emission from neutral, He- and H-like iron. A relativistically broadened component of the line was never included in our fits. We will discuss its possible contribution to the EW of the narrow core in Sect. \ref{discussion}. In the following, errors correspond to the 90\% confidence level for one interesting parameter ($\Delta \chi^2 =2.71$), where not otherwise stated. The adopted cosmological parameters  are $H_0=70$ km s$^{-1}$ Mpc$^{-1}$, $\Lambda_0=0.73$ and $q_0=0$ (i.e. the default ones in \textsc {Xspec} 12.2.1).

\section{The IT effect}

\begin{figure*}
\begin{center}
\epsfig{file=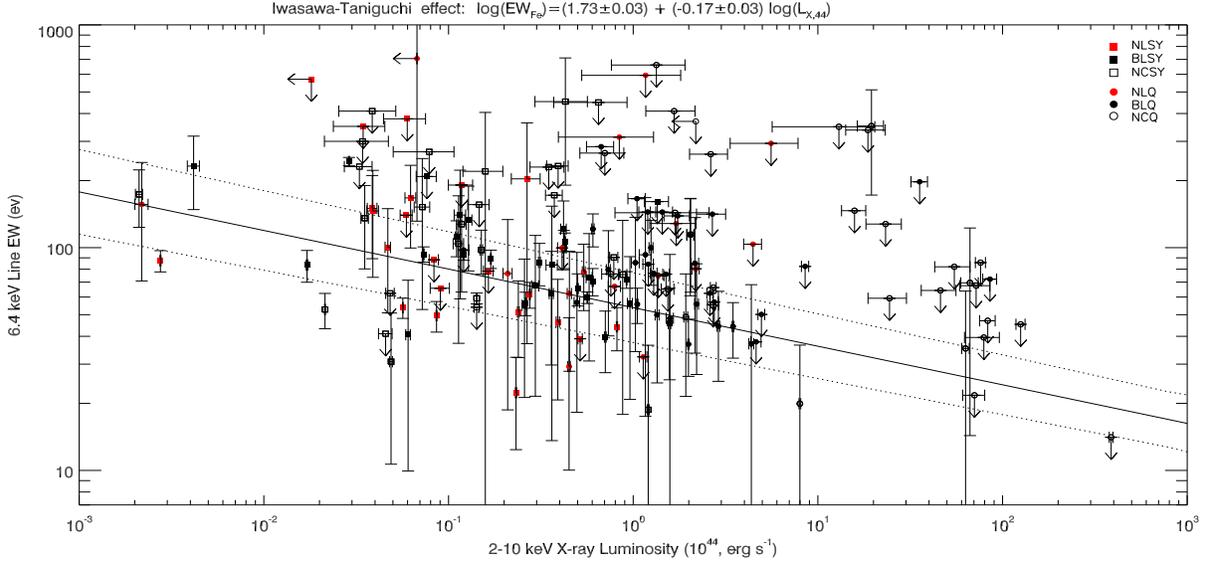, width=16cm, height=7.5cm}
\end{center}
\caption{\label{iwasawa}The `IT effect': neutral iron EW against 2-10 keV X-ray luminosity of the objects in our catalogue. The anti-correlation between the two parameters is shown as the best fit line, whose analytical expression is reported on the top. The broken lines represent the combined error on the slope and normalization of the best fit. The different symbols refers to the classification of the objects, on the basis of their absolute magnitude and H$\beta$ FWHM: \textit{NLSY}, narrow-line Seyfert 1; \textit{BLSY}, broad-line Seyfert 1; \textit{NCSY}, not-classified Seyfert 1 (no H$\beta$ FWHM measure available); \textit{NLQ}, narrow-line quasar; \textit{BLQ}, broad-line quasar; \textit{NCQ}, not-classified quasar (no H$\beta$ FWHM measure available). See text for details.}
\end{figure*}

Fig. \ref{iwasawa} shows the IT effect in our data. As a result of the fitting procedure described in Appendix \ref{appa}, we obtained the following best fit (see also Table \ref{fit}):

\begin{equation}
\label{iteffect}
\log(EW_{Fe}) = \left(1.73\pm0.03\right) + \left(-0.17\pm0.03\right) \log(L_{X,44})
\end{equation} 

\noindent where $EW_{Fe}$ is the EW of the neutral iron K$\alpha$ line in eV and $L_{X,44}$ is the 2-10 keV X-ray luminosity in units of $10^{44}$ erg s$^{-1}$. The Spearman's rank coefficient for this correlation is -0.33, corresponding to a Null Hypothesis Probability of $4\times10^{-5}$.

\begin{table}
\caption{\label{fit}Best fit parameters for the IT effect on the catalogue presented in this paper. The first line, in bold face, refers to the entire catalogue, while the following lines refers to Seyfert galaxies (S), Quasars (Q), Broad-line objects (B) and Narrow-line objects (N). Correlations with Eddington ratio (LD/C: luminosity-dependent or constant X-ray bolometric correction), BH mass and redshift follow.}
\begin{center}
\begin{tabular}{lllllllll}
\textbf{T} & \textbf{M} & \textbf{U} & \textbf{A} & $\mathbf{\sigma_A}$ & \textbf{B} & $\mathbf{\sigma_B}$ & $\mathbf{\rho}$ & \textbf{P} \\ 
(1) & (2) & (3) & (4) & (5) & (6) & (7) & (8) & (9) \\
× & × & × & × & × & × & × & × & × \\
\hline
\multicolumn{9}{c}{IT effect: $\log(EW) = A + B\log(L_{X,44})$}\\
\hline
× & × & × & × & × & × & × & × & × \\
\textbf{157} & \textbf{81} & \textbf{76} & \textbf{1.73} & \textbf{0.03} & \textbf{-0.17} & \textbf{0.03} & \textbf{-0.33} & $\mathbf{4\times10^{-5}}$ \\ 
× & × & × & × & × & × & × & × & × \\
79 (S) & 53 & 26 & 1.73 & 0.06 & -0.16 & 0.06 & -0.25 & 0.04 \\
78 (Q) & 28 & 50 & 1.75 & 0.06 & -0.19 & 0.06 & -0.28 & 0.02 \\ 
62 (B) & 43 & 19 & 1.72 & 0.05 & -0.21 & 0.06 & -0.42 & 0.001 \\
38 (N) & 19 & 19 & 1.63 & 0.09 & -0.21 & 0.08 & -0.34 & 0.07 \\
× & × & × & × & × & × & × & × & × \\
\hline
\multicolumn{9}{c}{EW vs Eddington ratio: $\log(EW) = A + B\log(L_{bol}/L_{edd})$}\\
\hline
× & × & × & × & × & × & × & × & × \\
82 (LD) & 50 & 32 & 1.61 & 0.05 & -0.19 & 0.05 & -0.38 & $6\times10^{-4}$ \\
82 (C) & 50 & 32 & 1.64 & 0.05 & -0.16 & 0.06 & -0.26 & 0.03 \\
× & × & × & × & × & × & × & × & × \\
\hline
\multicolumn{9}{c}{EW vs BH mass: $\log(EW) = A + B\log(M_{BH,8})$}\\
\hline
× & × & × & × & × & × & × & × & × \\
82 & 50 & 32 & 1.73 & 0.04 & -0.07 & 0.04 & -0.15 & 0.3 \\
× & × & × & × & × & × & × & × & × \\
\hline
\multicolumn{9}{c}{EW vs z: $\log(EW) = A + B\log(z)$}\\
\hline
× & × & × & × & × & × & × & × & × \\
157 & 81 & 76 & 1.63 & 0.06 & -0.13 & 0.04 & -0.16 & 0.08 \\
× & × & × & × & × & × & × & × & × \\
\hline
\multicolumn{9}{c}{Ratio$_{Fe}$ vs X-ray luminosity: $\log(Ratio_{Fe}) = A + B\log(L_{X,44})$}\\
\hline
× & × & × & × & × & × & × & × & × \\
81 & 52 & 29 & 0.12 & 0.06 & 0.09 & 0.06 & 0.12 & 0.3 \\
× & × & × & × & × & × & × & × & × \\
\end{tabular}
\end{center}

Notes: Col. (1)-(2)-(3) Total number of points, measures and y upper limits; Col. (4)-(5)-(6)-(7) Best fit parameter and errors for the linear fit $\log(y) = A + B\log(x)$; Col. (8)-(9) Mean Spearman's rank coefficient and Null Hypothesis probability. See Appendix \ref{appa} for details.
\end{table}

Our result is consistent with previous claim for the IT effect, whether radio-loud objects were included or not. In the original paper by \citet{it93}, the slope of their anti-correlation was $-0.20\pm0.03$. The following results are all consistent with each other: $-0.17\pm0.08$ \citep{page04}, $-0.06\pm0.20$ \citep{jb05}. As for \citet{jww06}, they quote $-0.20\pm0.04$ for their whole sample and $-0.10\pm0.05$ for a sub-sample without radio-loud objects.

We also tested if this anti-correlation still holds for subclasses of our catalogue. As shown in Table \ref{fit}, there is no significant difference between Seyfert galaxies and quasars, nor between broad- and narrow-line objects, even if the lower statistics prevent us from drawing firm conclusions.

\begin{figure*}
\begin{center}
\epsfig{file=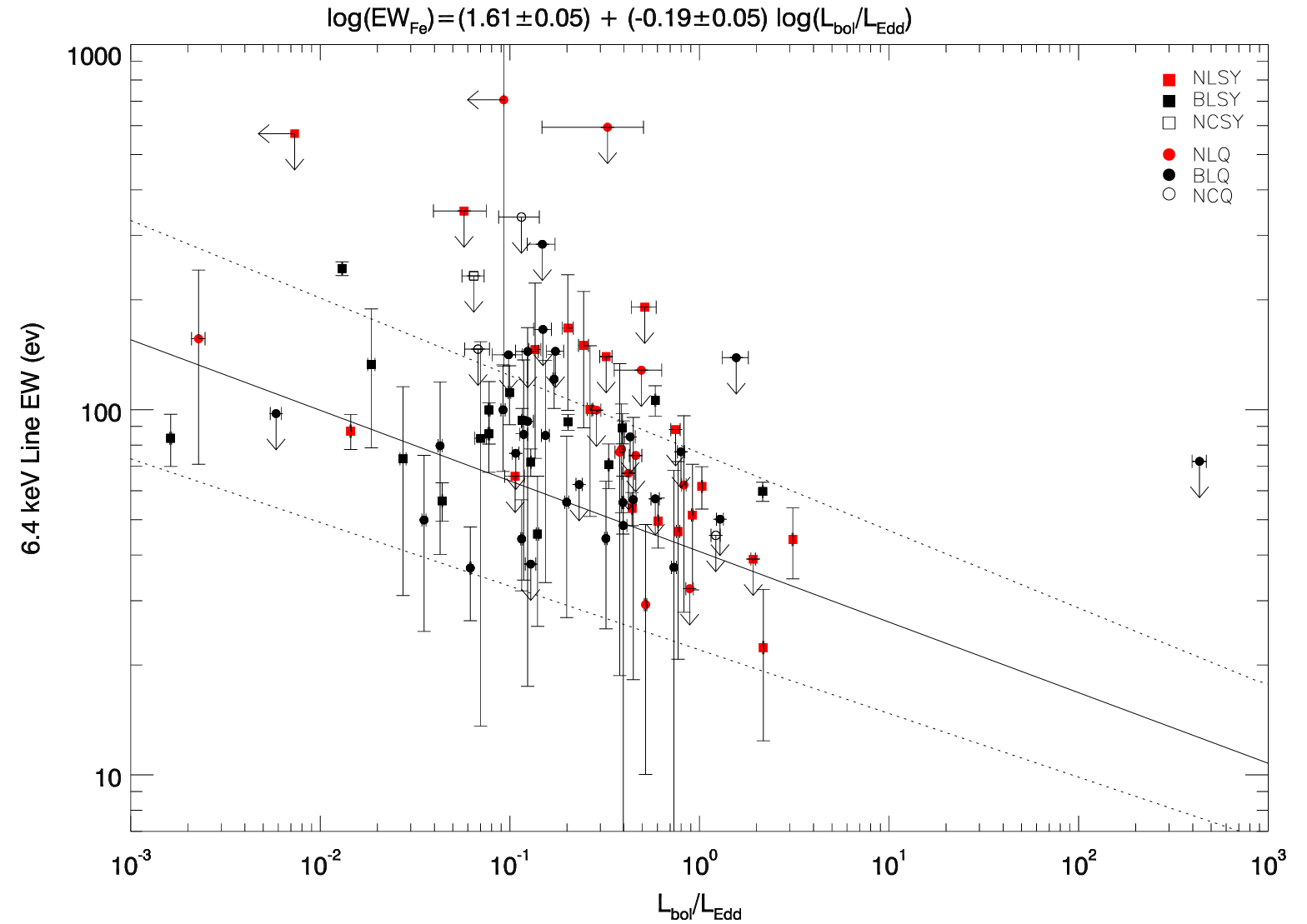, height=6cm}
\epsfig{file=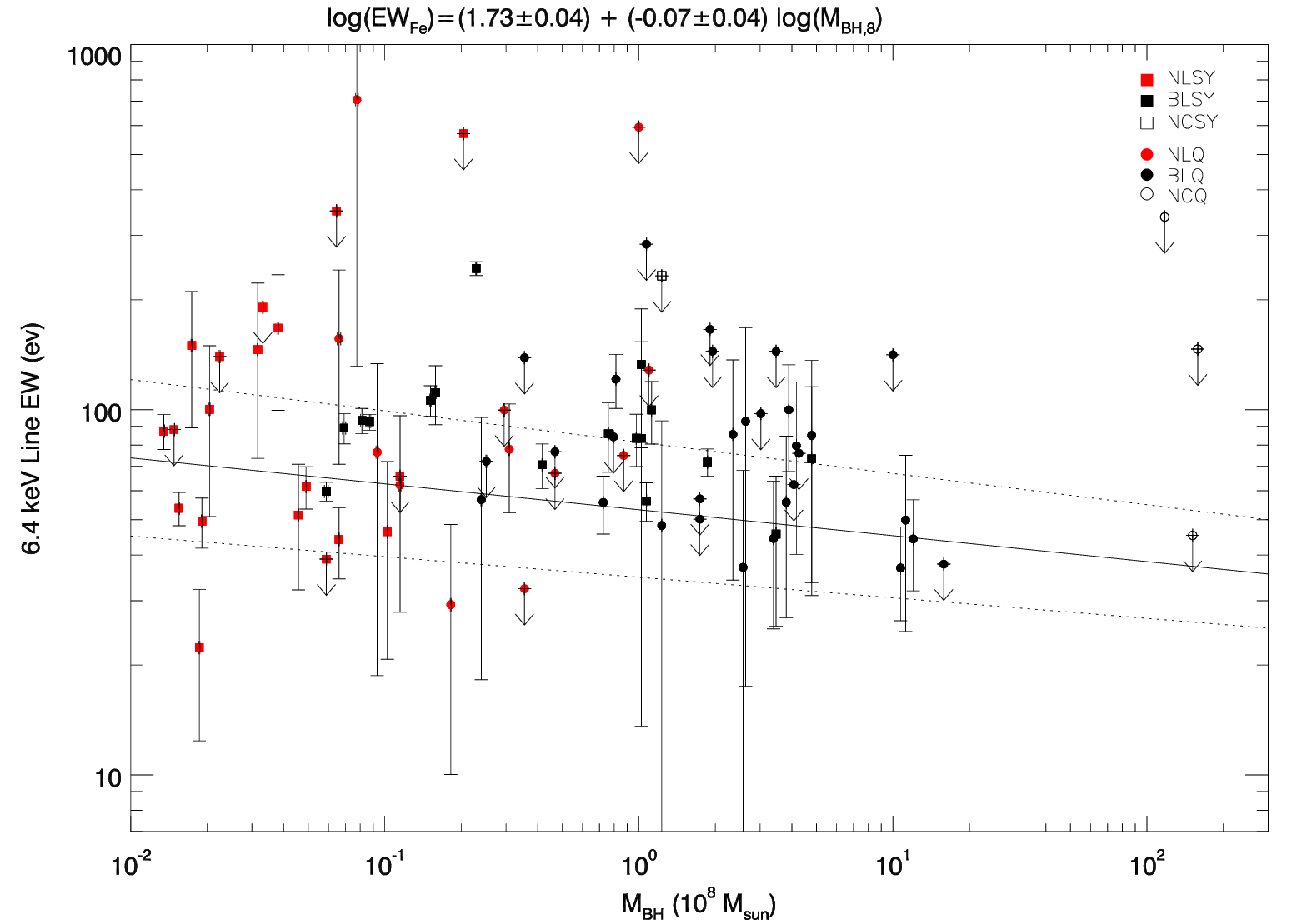, height=6cm}
\end{center}
\caption{\label{eddington}\textit{Left panel}: neutral iron EW against the Eddington ratio $L_{bol}/L_{edd}$, adopting a luminosity-dependent X-ray bolometric correction. \textit{Right panel}: The same as above, but against the BH mass. See caption of Fig. \ref{iwasawa} for details.}
\end{figure*}

In order to understand which is the main driver for the IT effect, we checked the behaviour of the iron K$\alpha$ EW with respect to the Eddington ratio, the BH mass and the redshift (see Table \ref{fit} and Fig. \ref{eddington}). The anti-correlation with the Eddington ratio is highly significant \citep[in particular when using a luminosity-dependent X-ray bolometric correction, like in][instead of a constant one]{mar04}. On the other hand, the correlation with z is weaker and, quite interestingly, no significant dependence of the iron EW on the BH mass is apparent.

\section{\label{discussion}Discussion}

We have shown in the previous section that our data confirm the significance of the IT effect. Several explanations have been proposed in the literature for this phenomenon.

The original Baldwin effect on the [{C\,\textsc{iv}}] $\lambda 1549$ \AA line was interpreted by \citet{mf84} in terms of a luminosity-dependent covering factor of the BLR. \citet{it93} noted that this explanation was also a good one for their recently discovered X-ray Baldwin effect, suggesting an origin of the iron K$\alpha$ line in a material close to the BLR itself. However, even if a contribution from the BLR cannot be excluded for most objects \citep[and in some cases seems to be the best possibility, see e.g.][]{bianchi03b}, no correlation exists between the width of Fe K and the H$\beta$ \citep{nan06}. Instead, the ubiquitous presence of an accompanying Compton reflection component strongly favours a scenario where most of the iron line flux comes from the torus \citep[e.g.][]{bianchi04}.

The IT effect should rather be associated to a change of the covering factor of a compact Compton-thick torus. Such a possibility could be related to disc-driven hydromagnetic wind models, which predict an increase of the opening angle of the torus with luminosity \citep{kk94}. This, in turns, would lead to a reduction of the covering factor of the torus, which could explain the observed iron line-luminosity anti-correlation. A side implication of this luminosity-dependent covering factor for the torus is a decrease of the fraction of obscured AGN with luminosity, which is indeed what recently observed \citep{ueda03,lafranca05}. It must be noted, however, that the latter observational result is based on Compton-thin sources only, while the torus is likely Compton-thick. 

Indeed, the decrease of obscured sources with luminosity can be explained by gravitational effects of the BH on the molecular gas in the disk of the host galaxy \citep{lamastra06}. These authors showed that the covering factor of this large-scale, Compton-thin material decreases with luminosity. Lamastra et al. (in prep.) investigated the iron line emission from the molecular gas in the disk, as expected from their model. As a natural consequence of their assumptions, the EW appears to be anti-correlated with the X-ray luminosity indirectly, the BH mass being the principal driving mechanism. However, Fig. \ref{eddington} (right panel) seems to indicate that no anti-correlation with the BH mass is observed. Moreover, due to the fact that the material is Compton-thin, the EW calculated by Lamastra et al. (in prep.) can only account for less than a half of the observed iron line EW, at each luminosity. Therefore, it seems unlikely that their model can explain the observed IT effect.

\begin{figure}
\begin{center}
\epsfig{file=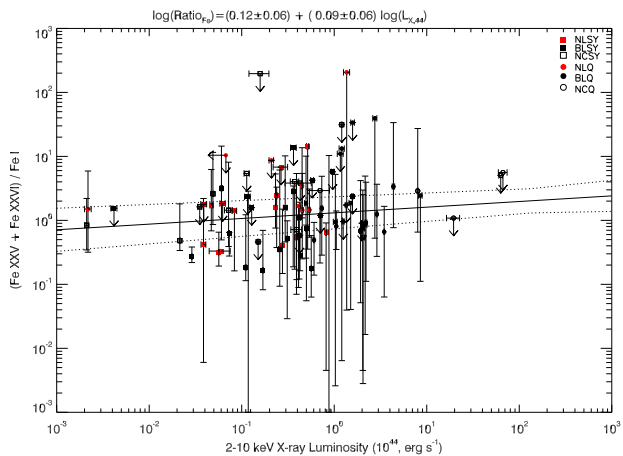, height=6cm}
\end{center}
\caption{\label{feionplots}The ratio between the ionized and the neutral iron lines against the 2-10 keV X-ray luminosity for the objects in our catalogue. See caption of Fig. \ref{iwasawa} for details.}
\end{figure}

Another possibility is that the ionization state of the iron emitting material changes according to luminosity. Even if \citet{nan97b} and \citet{nay00} proposed this scenario for the iron line component produced in the accretion disk, it is still possible that even the inner walls of the torus may become so ionized to suppress iron line emission for high-luminosity objects. As the ionization state rises, one would expect a strong contribution from He- and H-like iron \citep[e.g.][]{mbf96}. We tested this scenario in Fig. \ref{feionplots} (right panel), where the ratio of the overall contribution from highly ionized iron to neutral iron is plotted against X-ray luminosity. The best fit between these quantities suggests indeed a weak correlation, but its significance is poor (see also Table \ref{fit}). 

Recently, \citet{jww06} proposed that the IT effect can be completely accounted for by variability effects, assuming constant iron line fluxes. This naturally produces an anti-correlation between X-ray luminosity and iron EW, which may contribute to the overall IT effect. However, the simulated anti-correlation has a slope of $-0.05\pm0.05$ \citep{jww06}, much weaker than the observed IT effect. In any case, the amplitude of variability in radio-quiet AGN is generally smaller than one order of magnitude, while the IT effect is observed on six orders of magnitude in luminosity (see Fig. \ref{iwasawa}), making this explanation unlikely, as originally noted by \citet{it93}. 

A last comment must be spent on the possibility that the IT effect may be completely or partly due to the relativistically broadened component of the iron line. \citet{gbd06} showed that a broad profile is significantly more common in low luminosity AGN. Their stacked spectral residuals analysis on $\simeq100$ sources clearly presents a weaker red wing and narrow core for more luminous objects. Even though it is very difficult to quantitatively disentangle the contribution of each component to the overall EW decrease, due to the limited statistics of most objects, the EWs measured in our fits are likely dominated by the narrow core, especially when the underlying relativistic component is very broad. Therefore, the relativistic iron line can possibly contribute to the overall anti-correlation, but cannot account for most of the observed Iwasawa-Taniguchi effect, which implies an $\simeq80\%$ decrease of the iron EW in four orders of magnitude in X-ray luminosity (see our best fit in Eq. (\ref{iteffect})).

\acknowledgement
We would like to thank F. La Franca, A. Lamastra and G.C. Perola for useful discussions, C. Gordon for his help with \textsc{Xspec} and the anonymous referee for valuable comments. SB and GM acknowledge financial support from ASI (grant 1/023/05/0).

\begin{appendix}

\section{\label{appa}Linear fits on censored data}

Since we are dealing with a large number of upper limits for the EWs, we followed \citet{gua06} to perform a linear fit, taking into account both upper limits and errors on the y parameter. Their method is an extension of the regression method on censored data described by \citet{schm85} and \citet{isobe86} and can be summarized as follows. A large number of Ordinary Least Squares (OLS: Y vs. X) fits was performed on a set of Monte-Carlo simulated data derived from the experimental points according to the following rules: a) each detection was substituted by a random Gaussian distribution, whose mean is the best-fit measurement and whose standard deviation is its statistical uncertainty; b) each upper limit $U$ was substituted by a random uniform distribution in the interval [0,U]. The mean of the slopes and the intercepts derived from the fits of each data set are our `best fit'.
Note that errors and upper limits on the x parameter are not considered in the procedure, although plotted in Fig. \ref{iwasawa}, \ref{eddington} and \ref{feionplots}. This does not affect the results significantly, given the fact that they are typically much smaller than the errors on the EW. Moreover, the errors on the best fit parameters only take into account the experimental uncertainties, being the dispersion of these parameters within the Monte Carlo simulated data sets. Any intrinsic, physical scattering of the data points is not included. To test the significance of this anti-correlation, we calculated the Spearman's rank coefficient for each Monte Carlo simulated data set: the mean value is taken as the Spearman's rank coefficient of our `best fit'.

\end{appendix}

\bibliographystyle{aa}
\bibliography{sbs}

\end{document}